\documentclass[final,2p,times,twocolumn]{elsarticle}

\newcommand{\nn}{\nonumber}
\def\lesssim{\buildrel < \over {_{\sim}}}
\def\gtrsim{\buildrel > \over {_{\sim}}}

\begin{document}

\begin{frontmatter}

\title{Gamma Rays and Neutrinos from SNR RX J1713.7-3946}

\author[inaf]{G. Morlino}
\ead{morlino@arcetri.astro.it}

\author[inaf,infn]{P. Blasi}
\ead{blasi@arcetri.astro.it}

\author[inaf]{E. Amato}
\ead{amato@arcetri.astro.it}

\address[inaf]{INAF/Osservatorio Astrofisico di Arcetri, 
Largo E. Fermi, 5,50125 Firenze, Italy}

\address[infn]{INFN, Laboratori Nazionali del Gran Sasso, I-67010
 Assergi (AQ), Italy}

\begin{abstract}
The supernova paradigm for the origin of galactic cosmic rays can be
tested using multifrequency observations of both non-thermal and thermal
emission  from supernova remnants. The smoking gun of hadronic acceleration
in these sources can, however, only be provided by the detection of a high
energy neutrino signal. Here we apply the theory of non-linear particle
acceleration at supernova shocks to the case of the supernova remnant
RX~J1713.7-3946, which is becoming the stereotypical example of a possible
hadronic accelerator after the detection of high energy gamma rays by the
HESS telescope. Our aim is twofold: on one hand we want to address the
uncertainties in the discrimination between a hadronic and a leptonic
interpretation of the gamma ray emission, mainly related to the possibility
of a statistical uncertainty in the energy determination of the gamma ray
photons in the TeV region. On the other we want to stress how a $km^3$
neutrino telescope would break the degeneracy and provide evidence for
efficient cosmic ray acceleration in RX~J1713.7-3946. A $3\sigma$ evidence
would require about two years of observation. 
\end{abstract}

\begin{keyword}
acceleration of particles -- neutrinos -- object: RX~J1713.7-3946
\end{keyword}

\end{frontmatter}

\section{Introduction}
\label{sec:intro}

Supernova remnants (SNRs) are the most plausible sources of galactic
cosmic rays. The recent detection of multi-TeV gamma radiation from
several SNRs makes the case stronger, especially when coupled with
multifrequency observations of the same remnants. However, despite all
this progress, the evidence that SNRs are indeed the main contributors
to Galactic cosmic rays remains circumstantial. A smoking gun evidence
of efficient acceleration of cosmic rays in these sources can only
come from the detection of high energy neutrinos, resulting from the
decays of charged pions within the source. 

The acceleration of cosmic rays at supernova blast waves is well
described by the non linear theory of diffusive shock acceleration 
(NLDSA) (see \cite{maldrury} for a review). This theory allows us to 
calculate the spectrum and spatial distribution of cosmic rays
accelerated at a supernova shock taking into account the dynamical
reaction of the accelerated particles on the shock and, in its most 
recent version \cite[]{amato06,apjlett,long}, also the generation of
magnetic field through streaming instability induced by the
accelerated particles \cite[]{bell78,bell04} and the dynamical reaction 
of the amplified magnetic field on the plasma. Such a dramatic
improvement in the quality of our theoretical approach allows us to
finally compare theory with observations and make testable predictions 
for future observations.

One of the most important recent breakthroughs in establishing the SNR
paradigm for the origin of cosmic rays has been the detection of
narrow filaments of  non-thermal X-ray emission in the direction of supernova remnant
rims \cite[]{bamba03,bamba05, laz03, laz04, vink03}. The filaments are
most commonly interpreted as the result of severe synchrotron energy
losses of ultra relativistic electrons which are forced to radiate a
large fraction of their energy (in the form of X-rays) in a narrow
region downstream of the shock. The required strengths of the magnetic 
field are of order $\sim 100-500\mu G$ downstream of the shock. Such large levels  
of magnetization might be the manifestation of magnetic field amplification 
in the upstream region, as due to the streaming of
cosmic rays \cite[]{skillinga,bell78,amato06}. Not only the detection of
large magnetic fields signals for efficient cosmic ray acceleration,
but in turn large magnetic fields are needed for increasing the maximum
energy of accelerated particles to the knee region \cite[]{dammax}, namely 
for efficient acceleration. Efficient cosmic ray acceleration and magnetic field 
amplification are two sides of the same coin.

On the other hand, it has been proposed that the interpretation of the
narrow X-ray rims may be flawed \cite[]{pohl05}: they could in fact
result from damping of magnetic field in the downstream region. The
emission region would be narrow because the field disappears, not
because particles lose energy effectively. The situation is
currently subject of active debate: the damping would naively result
in the appearance of filaments not only in X-rays, but also in the
radio emission, and at present there is no evidence for such a
phenomenon. In addition, the absence of magnetic field amplification
(or a mitigation of the effect) would reduce the maximum energy
achieved by the accelerated particles, unless the shock configuration
is quasi-perpendicular \cite[]{jokipii87}. An additional possibility has
been proposed in \cite{giajo}: the magnetic field could be amplified
downstream (and not upstream) because of the development of
corrugations on the shock surface which result in eddies in which the
magnetic field winds up and gets amplified. In order for this scenario
to lead to large maximum momentum of the accelerated particles, the
field direction must be very inclined with respect to the shock normal 
in the upstream region. 

A powerful diagnostic tool for particle acceleration in SNRs is
represented by the gamma ray emission in the GeV-TeV energy region. 
The spectrum of the gamma ray emission, its extension to high energies
and the shape of the cutoff all provide precious information on
whether the radiation is of leptonic or hadronic origin. In
\cite{morlino08} we discussed in detail the application of NLDSA to the
case of the supernova remnant RX~J1713.7-3946. If the HESS data on this
remnant are taken at face value, then it is problematic to fit them
in the context of a leptonic model for at least two reasons: 1) the    
spectral cutoff expected in leptonic scenarios leads to a gamma ray   
emission which falls short of the highest energy data points measured
by the HESS telescope; 2) in any case, fitting both the TeV and X-ray data 
requires the presence of a very large background of infrared photons 
in the remnant. 

The second point raised above should be considered as a circumstantial
evidence against a leptonic model, but by itself it does not suffice
to rule out this class of models. In principle the first point is rather
solid, but experimental effects might weaken its significance: a
statistical (and systematic) uncertainty in the energy determination
affects the steep gamma ray spectrum by making it look smoother than it
actually is so that convolving the theoretical prediction with an estimate
of the uncertainty in the energy determination makes the leptonic spectrum
look more similar to the hadronic case. We address this issue in a
quantitative way here.

The hadronic interpretation appears to be more sound, but there are
shortcomings in this case as well: first, the thermal X-ray flux expected
from RX~J1713.7-3946  is larger than the observed radiation of synchrotron origin.
Second, the number of electrons needed to explain the observations is about
$10^{-4}$ of the number density of accelerated protons, at odds with the
e/p ratio observed at Earth at energy $\sim 1-10$ GeV. 

Both these points are however rather weak at the present time. The first
is based on the assumption that electrons and protons share the same
temperature downstream. This condition is hardly achievable and in fact
one can easily argue that the temperature of electrons should be much
smaller than that of protons. On the other hand, Coulomb scattering might
be sufficient to raise the electron temperature to a level large enough to
excite emission lines of heavy elements \cite{patnaude09}. The second
point is equally weak in that electrons might be accelerated at different
stages of the SNR. Moreover, recent data from PAMELA \cite{pam} and ATIC
\cite{atic} suggest that a substantial contribution to the observed
electron spectrum at Earth might come from sources other than SNRs \cite{hooper,stanev}. 

The safest way of proving or rejecting acceleration of hadrons in  RX~J1713.7-3946,
as well as in other remnants, is to search for  neutrinos produced in the
decays of charged pions. In this paper we apply the NLDSA model developed
by \cite{amato05,amato06}, and previously used to describe the
multifrequency spectrum of RX~J1713.7-3946 {} \cite[]{morlino08}, in order to calculate
the expected neutrino flux from this remnant. 

Previous attempts at estimating the neutrino flux from SNRs have been
typically based on phenomenological approaches, building on the assumption
of a power law approximation for the gamma ray spectrum and simple scaling
relations between the gamma ray and neutrino spectra. 

In \cite{alv02} the authors estimated a flux of 40 neutrino induced muons
(and antimuons) per $km^2$ per year from RX~J1713.7-3946, using a power law for the
$\gamma$-ray spectrum $f_\gamma(E) \propto E^{-2}$, based on CANGAROO
observations; the maximum neutrino energy was assumed to be $\sim 10$ TeV.
Neutrino oscillations, absorption and location of the detector were not
taken into account. Such effects were included by \cite{cos05}, where
$f_\gamma(E) \propto E^{-2.2}$ was used, the power law index being
inferred from HESS data (no cut-off energy was assumed in this
approach). The authors obtain a flux of neutrino induced muons
$N_{\mu+\bar\mu} \simeq 10\ {\rm km^{-2}\ yr^{-1}}$. Again based on HESS 
data, \cite{vis06} used two different parametrizations for the $\gamma$-ray
spectra, a power law plus an exponential cutoff at $E_{\gamma, {\rm max}}=
12$ TeV and a broken power law (with a knee at 6.7 TeV), leading to
$N_{\mu+\bar\mu}= 4.8$ and 5.4 ${\rm km^{-2}\ yr^{-1}}$ respectively. An
attempt to extract the proton spectrum from the HESS data on the gamma ray
flux (assumed to be of hadronic origin), and compute the neutrino
flux based on that, was done in \cite{vil07}.

In the present paper we carry out the calculations by using our model of
NLDSA which provides a self-consistent description of the acceleration 
of cosmic rays in the remnant. This leads to a shape of the gamma ray 
spectrum which is not a simple power law, due to the non linear
effects induced by the dynamical reaction of the accelerated particles and
the amplified magnetic field. At the same time we also obtain
self-consistently the neutrino spectrum, and we use it to derive the
expected number of events in a $km^3$ neutrino telescope. 

The paper is organized as follows: in \S~\ref{sec:theory} we summarize
the technical aspects of NLDSA and its application to particle
acceleration in supernova remnants. We also describe the calculations
of the non thermal radiation from RX~J1713.7-3946, with special attention for the
gamma ray emission in both the hadronic and leptonic scenario. We
discuss in detail the possibility to use present and future observations to
discriminate between the two, once a statistical uncertainty in the energy
determination of the photon events is taken into account. We show that a
leptonic scenario convolved with a gaussian distribution of the photon
energies with $\delta E/E\sim 30\%$ or larger leads to the impossibility to
distinguish the leptonic prediction from the hadronic one, at least using
the HESS data on RX~J1713.7-3946. The implications for future gamma ray
telescopes are also briefly discussed. On these premises it is very
important to aim at the detection of the associated neutrino signal, whose
intensity is calculated in \S~\ref{sec:signal}, where we also estimate the
neutrino induced muon signal in a $km^3$ neutrino detector. We conclude in
\S~\ref{sec:conc}. 

\section{NLDSA and the non thermal emission of RX~J1713.7-3946}
\label{sec:theory}

\subsection{Spectrum of accelerated protons and electrons}
\label{sec:max_energy}

In NLDSA theory, the overall shock structure and the outcome of the
particle acceleration process are inextricably linked. When
acceleration is efficient, the pressure of accelerated particles affects 
the shock dynamics, leading to the formation of a precursor, 
namely a region where the fluid velocity progressively decreases 
while approaching the shock from far upstream. At the same time the
streaming of accelerated particles is responsible for the
instabilities that lead to magnetic field amplification. In turn, the
fluid profile in the precursor and the amplified, turbulent magnetic
field, with the scattering it provides (and possibly the induced
energy losses), determine the efficiency of particle acceleration and the
resulting spectrum, including its high energy cutoff.

The shock structure and the accelerated particle spectrum are 
computed as in \cite{morlino08}: the basic structure of the
calculation is the same proposed in \cite{amato05} and \cite{amato06}, 
but with a crucial new aspect taken into account, namely the dynamical 
reaction of the self-generated magnetic field, which is included
following the treatment of \cite{apjlett} and \cite{long}. This means
that the conservation equations at the shock and in the precursor are
modified so as to include the magnetic contribution. The compression
factor at the subshock, $R_{sub}$, and the total compression factor,
$R_{tot}$, are deeply affected by this change, resulting in a decrease
of the compression ratio in the precursor, $R_{tot}/R_{sub}$, as soon
as the amplified magnetic field contributes a pressure comparable to
that of the thermal gas upstream. This smoothening of the precursor
reflects in spectra of accelerated particles which are closer to power
laws, though the concavity typical of NLDSA remains visible \cite[]{long}. 

The normalization of the proton spectrum is an output of our non linear
calculation, once a recipe for injection has been established. Following
\cite{bgv05}, particles are injected immediately downstream of the 
subshock. The fraction of particles crossing the shock surface which are
injected in the acceleration process, $\eta_{\rm inj}$, can be written as
\begin{equation}
\eta_{\rm inj}= 4/\left(3\sqrt{\pi}\right) 
(R_{\rm sub}-1) \, \xi^3\,e^{-\xi^2}\,.
\label{eq:inj}
\end{equation} 
Here $\xi\sim 2-4$ is defined by the relation $p_{\rm inj}=\xi \, p_{\rm
th,2}$, where $p_{\rm th,2}$ is the momentum of thermal particles
downstream. $\xi$ parametrizes the poorly known microphysics of the
injection process, but $p_{\rm th,2}$ is an output of the problem: as
a result, the injection efficiency is affected by the dynamical
reaction exerted by the accelerated particles and by the amplified
magnetic field. 

Finally, the maximum momentum of the accelerated particles is determined
following \cite{dammax} for the computation of the acceleration time
in the presence of a precursor. 

For protons we use as a prescription the equality between the acceleration
time and the age of the SNR:
\begin{equation}
t_{acc}(p_{p,\rm max})=t_{SNR}\ .
\label{eq:pmax}
\end{equation}

For electrons, energy losses can be important. Their maximum momentum
$p_{e,\rm max}$ is determined by equating the acceleration time with
the minimum between the time for energy losses and the age of the
remnant. The loss time of electrons over a cycle of shock crossing
needs to be weighed by the residence times, $t_r$, upstream and
downstream, so that the condition for the maximum momentum, in the
loss dominated case, can be written as:
\begin{equation}
t_{acc}(p) =
\frac{t_{r,1}(p)+t_{r,2}(p)}{{\frac{t_{r,1}(p)}{\tau_{l,1}(B_1,p)}}+
{\frac{t_{r,2}(p)}{\tau_{l,2}(B_2,p)}}}
\label{eq:pemax}
\end{equation}
where $\tau_l$ denotes the loss time, and the indexes ``1'' and ``2''
refer to quantities measured upstream and downstream respectively. The
residence times in the context of the non linear theory of particle
acceleration can be written explicitely (from Eqs.~(25) and (26) of
\cite{dammax}). Eq.~(\ref{eq:pemax}) must be solved numerically for
$p_{e,\rm max}$, contrary to the case of acceleration in the test particle
regime. However an approximate analytical solution, valid when only
synchrotron losses are important, was also proposed in \cite{morlino08}.

As to the electron spectrum at the shock, $f_{e,0}(p)$, this is easy
to calculate for $p\ll p_{e,\rm max}$. In fact, at a given momentum
$p$, the slope of the electron and proton spectrum is the same, if one
assumes that both species experience the same diffusion coefficient. What
is unconstrained {\it a priori} is the relative normalization of the two
spectra, $K_{ep}$, which can only be obtained by fitting the observations.
This is reasonable since electrons do not exert any appreciable dynamical
reaction on the shock.

The spectrum of electrons at energies around and above $p_{e,\rm max}$,
namely the shape of the cutoff is harder to calculate in the context
of non-linear theory. Since the spectra we find for electrons at
$p<p_{e,\rm max}$ are not far from being power laws with slope $\sim
4$, we adopt the modification factor calculated by \cite{zir07} for
strong shocks in test particle regime. The resulting electron spectrum
at the shock, in the loss dominated case, is:
\begin{equation} \label{eq:f_e(p)}
f_{e,0}(p)= K_{ep}\,f_{p,0}(p)
 {\left[1+0.523 \left(p/p_{e,\rm max}\right)^{\frac{9}{4}}\right]}^2
 e^{-p^2/p_{e,\rm max}^2} \,.
\end{equation} 
What is important to notice in this expression is that the cutoff is
not a simple exponential, a fact which reflects in the shape of the 
synchrotron spectrum radiated by the electrons, making it different
from what assumed by most of the previous work on the subject. On the 
other hand, if the maximum momentum is indeed determined by the age of the
remnant, then the cutoff shape is expected to be exponential.

\subsection{Magnetic field amplification and compression} 
\label{sec:dyn}

The turbulent magnetic field close to the shock can be enhanced by
several physical processes. However, here, as in \cite{morlino08}, we 
focus on the amplification due to resonantly excited streaming
instability induced by cosmic rays accelerated at the shock. Resonant
streaming instability \cite[]{skillinga} is likely responsible for
most of the magnetic field amplification in SNRs after the beginning
of the Sedov phase \cite[]{kinetic}, while the non-resonant mode of the
same instability \cite[]{bell04} is more effective at earlier times.
When the predictions of linear theory are extrapolated to the
non-linear regime of field amplification (which one is forced to do
for lack of a better treatment), the resulting field strengths are in 
agreement with the values inferred by identifying the thickness of
the X-ray filaments with the synchrotron loss length of the highest energy
electrons. The strength of the magnetic field at the position $x$ upstream, $\delta
B(x)$, in the absence of damping, can be estimated from the saturation
condition, that, for modified shocks, reads \cite[]{long}:
\begin{equation} \label{eq:B_amp}
p_w(x) = U(x)^{-3/2} \left[\frac{1-U(x)^2}{4\,M_{A,0}} 
           \right] \,,
\end{equation}
where $p_w(x)= \delta B(x)^2/(8\pi \rho_0 u_0^2)$ is the magnetic
pressure normalized to the incoming momentum flux at upstream infinity,
$U(x)=u(x)/u_0$ and $M_{A,0}=u_0/v_A$ with $v_A$ the Alfv\'en
velocity at upstream infinity, where only the background magnetic
field, $B_0$, assumed parallel to the shock normal, is present. 
Eq.~(\ref{eq:B_amp}) correctly describes the effect of compression 
in the shock precursor through the term $U(x)^{-3/2}$. For the
upstream temperature that we adopt in RX~J1713.7-3946{} (see below),
damping in the upstream region is expected to be negligible.

The magnetic field downstream of the subshock is further enhanced by
compression, according to:
\begin{equation} \label{eq:B_2}
B_2= R_{sub} \, B_1,
\end{equation} 
where $B_1$ is the magnetic field immediately upstream of the subshock
and we have used the fact that we are dealing with Alfv\'en waves, and
hence turbulence perpendicular to the shock normal. 

\subsection{Computation of the radiation fluxes and spectra}
\label{sec:rad}
The flux of non-thermal radiation at $\gamma$-ray photon energies is
computed as in \cite{morlino08}. In that work we considered both 
possible scenarios for the origin of high energy photons in 
RX~J1713.7-3946, namely a leptonic origin, through inverse Compton scattering
(ICS) of ambient low energy photons by accelerated electrons, or a
hadronic origin, from $\pi^0$ decay following nuclear collisions of 
relativistic protons.

In the hadronic scenario, the $\gamma$-ray flux from RX~J1713.7-3946{} is accompanied
by a flux of neutrinos coming from the decay of charged pions that are
produced in nuclear collisions together with neutral pions. We compute the
fluxes and spectra in both channels using the approximated expressions for
the cross sections as provided in Ref. \cite{kel06}.

The $\gamma$-ray and the neutrino fluxes, $\Phi_i^0(E_i)$ ($i=\gamma,\nu$),
produced by \textit{p-p} collisions from a source located at distance $d$
from Earth, can be expressed as follows:
\begin{equation} \label{eq:flux}
\Phi_i^0(E_i)= \frac{c}{4\pi d^2} \int d\textbf{r} \,n(\textbf{r})
    \int_{E_i}^\infty dE_p f_p(\textbf{r},E_p)
    \frac{d\sigma_i(E_p,E_i)}{dE_i} \,.
\end{equation}
In this expression, the apex $0$ is used to indicate the neutrino flux
that would have been seen at the Earth in the absence of neutrino
oscillations. For gamma rays the two fluxes clearly coincide. Here $n$ is
the gas (target) density in the SNR and $f_p(\textbf{r},E_p)$ is the
distribution function of accelerated protons at the location $\textbf{r}$
at a given energy $E_p$. In general both quantities depend on the location
in the shell, but  here we assume them to be constant in the region
enclosed between the contact discontinuity and the forward shock and
vanishing outside this region. Finally, $d\sigma_i/dE_i$ is the inclusive
differential cross section for the production of particles of type 
$i$. This quantity is usually expressed through the total inelastic
$p$-$p$ cross section, $\sigma_{\rm inel}$, and the dimensionless
distribution function $F_i(x_i,E_p)$ for secondaries, as
\begin{equation} \label{eq:cross_sec}
 \frac{d\sigma_i(E_p,E_i)}{dE_i} = \frac{\sigma_{\rm inel}(E_p)}{E_p}
     \, F_i\left(x_i, E_p \right) \,,
\end{equation}
where $x_i \equiv E_i/E_p$ is the fraction of proton energy
transferred to the secondary particle. 
For the functions $F_i$, which effectively enclose all the details of
the hadronic processes involved, we use the
analytical approximation derived in \cite{kel06} on the basis of
numerical simulations of $p$-$p$ collisions with the publicly
available code SIBYLL. The analytical formulae provide a very good
description of the flux and energy distribution of secondaries for 
energies above 0.1 TeV. As far as photons are concerned, $F_\gamma$ 
also includes the contribution of $\eta$ meson decay, in addition to 
that of $\pi^0$, with an overall accuracy of order a few~\%. The 
estimate of neutrino fluxes is slightly less accurate, because $F_\nu$ 
only includes the decay of charged pions, while neglecting the
contribution from $K$-mesons, and therefore leading to underestimate 
the neutrino flux by about 10~\%.
At energies lower than 100 GeV, and down to the rest energy of the 
$\pi$-meson, we use the extrapolated formulae provided again by
\cite{kel06}, that should be accurate within 10~\%. 

The flux and spectrum of ICS photons is calculated by using the exact
kernel for ICS, with the full Klein-Nishina (KN) cross section
\cite[]{rybicki}. The main target photon field contributing to emission in
the high energy $\gamma$-ray band is provided by dust-processed infrared 
photons, described by a blackbody spectrum with temperature $\sim 20$
K \cite[]{sch98}. Following the instance of the IR+Optical photon background in
the interstellar medium (ISM) we assume that the ratio of the optical to
infrared energy densities remains $\sim 20$, while the energy density of IR
light, $W_{IR}$, is left as a free parameter (in the ISM, $W_{IR}\approx
0.05\, \rm eV\ cm^{-3}$).

In order to compute the spectra of accelerated particles and the
resulting emission, we need estimates for a number of environmental
parameters relative to RX~J1713.7-3946, namely its distance ($d$), age ($t_{SNR}$),
expansion velocity ($u_0$), and the values of temperature ($T_0$), density
($n_0$), and magnetic field strength ($B_0$), in the surrounding medium. In
addition to these, the only free parameters of the model are: $\xi$,
entering Eq.~(\ref{eq:inj}), and $K_{ep}$, defining the ratio between
accelerated electrons and protons. In the ICS scenario an additional free
parameter is the above mentioned $W_{IR}$. 

The uncertainties on the various parameters and how they affect the
results of our calculations are thoroughly discussed in
\cite{morlino08}. Here we only summarize the values  that we have
found to provide the best fit to the multifrequency data in
both scenarios. Our adopted values of distance, age, expansion
velocity and temperature are: $d=1\ {\rm kpc}$, $t_{SNR}=1600\ {\rm
 yr}$ (consistent with the historical chinese record of a supernova
explosion in AD 393 \cite[]{wang}), $u_0=4300\ {\rm km/s}$, $T_0=10^6\
{\rm K}$. Other parameters are, in the hadronic scenario: $n_0=0.12\
{\rm cm}^{-3}$, $B_0=2.6\ \mu G$, $\xi=3.8$; in the leptonic scenario,
instead:  $n_0=0.01\ {\rm cm}^{-3}$, $B_0=1.5\ \mu G$, $\xi=4.1$. 

The resulting values of magnetic field strengths, acceleration
efficiency and maximum energy of the accelerated 
particles are quite different in the two cases: the hadronic scenario
entails efficient acceleration, with a fraction of accelerated
particles corresponding to about $10^{-4}$ and a maximum proton energy
exceeding $10^{14}\ {\rm eV}$. In the leptonic case the fraction of
accelerated protons drops to less than $10^{-5}$, corresponding to
an energy conversion efficiency of about 2~\%. The magnetic field
downstream is of order 20 $\mu G$, to be compared with $B_2 \sim 100\
\mu G$ for the first scenario (in agreement with the value inferred from the 
synchrotron loss length interpretation of the rim thickness). 
The ratio between the number density of accelerated electrons
and protons at the shock turns out to be $K_{ep}\sim 10^{-4}$ if
acceleration is efficient. On the other hand, inefficient acceleration, and
the lower value of the magnetic field associated with the leptonic
scenario, would favour $K_{ep}\sim 10^{-2}$, in agreement with measurements
of the diffuse galactic cosmic rays. However, in order to fit both the
X-ray and the gamma-ray fluxes in the context of this purely leptonic
scenario a local energy density of infrared radiation $\sim 24$ times
larger than the galactic average is required. 

\begin{figure}
\begin{center}
\includegraphics[angle=0,scale=.9]{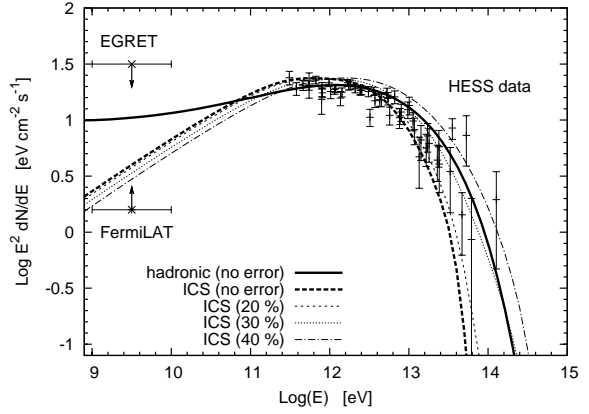}
\caption{Hadronic versus leptonic scenarios for the origin of TeV
 emission from RX~J1713.7-3946. In both cases, the flux is computed
 for the best-fit values of the parameters, as specified in the text.
 The thick solid line represents the spatially integrated spectral
 energy distribution of photons from $\pi^0$ decay in the hadronic
 scenario, while the thin solid line refers to ICS in the leptonic
 scenario. Symbols represent all available HESS data and also plotted 
 are EGRET upper limit and FermiLAT sensitivity for GeV energy photons
 from this source. Other lines on the plot represent the result of 
 convolution of the thin solid line with gaussians of different
 widths as explained in the text.} 
\label{fig:f1}
\end{center}
\end{figure}

In Fig.~\ref{fig:f1} we plot the gamma ray flux as obtained through our
calculations, for both a hadronic model (thick solid line) and a leptonic
one (thick dashed line) (see \cite{morlino08}). Taken at face
value, the curves in Fig.~\ref{fig:f1} clearly show that the hadronic
scenario reproduces the high energy observations much better, in
particular the highest energy data points of HESS. On the other hand,
the possibility to discriminate between the two models relies upon the
assumption that the statistical {\it uncertainty} in the energy
determination is sufficiently small. In order to address this point, we
calculate the theoretical prediction for the leptonic scenario in the case
in which there is a statistical uncertainty in the energy determination
$\sigma(E)=\chi E$ with $\chi=0.2,0.3,0.4$. Our results are shown again in 
Fig.~\ref{fig:f1} (different line-types are labelled in the figure). 

The thin curves in the figure are obtained by convolving the predicted
ICS flux at a given energy, $\phi_{\rm ICS}(E)$, with a gaussian of
given width, namely :
\begin{equation}
\Phi_{\rm ICS}'(E)=\int\ dE' \frac{\Phi_{\rm ICS}(E')}{\sqrt{2 \pi\
\sigma^2(E)}}\
\exp{\left[-\frac{(E-E')^2}{2 \sigma^2(E)}\right]}.
\label{eq:conv}
\end{equation}  

The conclusion that these curves lead to is that a statistical uncertainty
larger than $30\%$ in the energy region above 10 TeV would inhibit the
discrimination between the two models, based on HESS data alone. 
The nominal energy resolution of HESS telescope is around 15\%
\cite[]{aha08}, sufficient to allow for such discrimination. The
main issue in this case is the data statistics: the main differences
between the two scenarios, and the superior quality of the fit
obtained within the hadronic one, show at the highest energy, where
the statistical significance of the data points is not very high, as
shown by the large error bars in Fig.~\ref{fig:f1}. The role of the
energy resolution is however important also in view of future
gamma-ray telescopes, like CTA, whose design is currently being
discussed by the scientific community.

In addition to all this, it is worth keeping in mind that the Fermi
satellite is expected soon to provide another crucial bit of information 
to this debate, in that the level of detection (or the non detection) should
clarify the issue of a leptonic or hadronic origin for the gamma ray
emission from RX~J1713.7-3946, independent of the shape of the high
energy cutoff (see the low energy part of Fig.~\ref{fig:f1}).

\section{Neutrino signal}
\label{sec:signal}

The computation of the neutrino flux in the absence of neutrino
oscillations was carried out in the assumption of perfect isospin symmetry
($\Phi_{\pi^0} \simeq \Phi_{\pi^+} \simeq \Phi_{\pi^-}$), which leads to equal
neutrino and antineutrino fluxes of a given flavor. For our purposes this is a
good approximation.

The neutrino flux at the Earth is related to $\Phi_\nu^0$ through the
oscillation probabilities:
\begin{eqnarray} \label{eq:sum}
\Phi_{\nu_l}(E,d)= 
     \sum_{l'=e,\mu,\tau} P_{l l'}(E,d) \, \Phi_{\nu_{l'}}^0(E) \,.
\end{eqnarray}

The transition probabilities, $P_{l l'}$, are in general functions of 
energy and travel length, but since at the energies considered here
the oscillation lengths are very short compared with the typical size of
a SNR, the oscillation probability can be space averaged. The resulting
flux of $\nu_\mu$ crossing the Earth, not taking into account 
absorption, is: 
\begin{eqnarray} \label{eq:oscillation}
\Phi_{\nu_{\mu}}= P_{\mu\mu} \Phi_{\nu_{\mu}}^0 + P_{e\mu}
                  \Phi_{\nu_{e}}^0
                = 0.4\, \Phi_{\nu_{\mu}}^0 + 0.2\, \Phi_{\nu_{e}}^0 \,,
\end{eqnarray}
with an identical equation holding for antineutrinos. The errors due 
to the uncertainties in the oscillation parameters are negligible 
($\sim 5\%$). Since at the source $\nu_\mu$ and $\nu_e$ are produced 
in a ratio $\{2:1\}$, the effect of oscillations translates into a
flux of muon neutrinos at Earth that is $\sim 50\%$ of that produced at
the source: $\Phi_{\nu_{\mu}}=0.5\Phi_{\nu_{\mu}}^0$.

\begin{figure}
\begin{center}
\includegraphics[angle=0,scale=1]{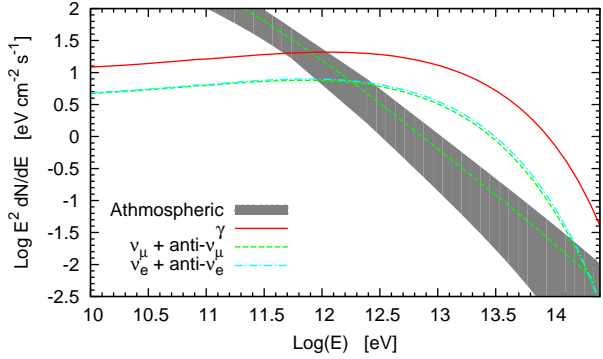}
\caption{Fluxes of photons and neutrinos (plus anti-neutrinos) from
RX~J1713.7-3946 predicted in the hadronic scenario. Both electronic and
muonic neutrinos contributions are shown. The shaded region represents the
range of theoretical predictions for the atmospheric muonic neutrino (plus
anti-neutrino) flux integrated in a cone of semi-aperture $0.5^\circ$.}
\label{fig:f2}
\end{center}
\end{figure}

The flux of neutrinos and antineutrinos of each flavor expected at Earth is
shown in Fig.~\ref{fig:f2}, and compared with the flux of gamma rays (solid
line). The flux of neutrinos has to be compared with the background whose
main contribution comes from atmospheric neutrinos. The shaded region
shown in Fig.~\ref{fig:f2} refers to the theoretical prediction for the
atmospheric neutrino flux as we explain below. 

Following the estimates of \cite[]{vol80,hon95,agr96} the atmospheric
neutrino flux above 1 TeV can be approximated as:
\begin{equation} \label{eq:nu_atm}
\Phi_{\nu_\mu+\bar\nu_\mu}^{\rm atm}(E_\nu) \simeq 
  4.6 \times 10^{-8} \left( \frac{E_\nu}{1 \,{\rm TeV}}\right)^{-3.7} 
  {\rm TeV^{-1} \, cm^{-2} \,s^{-1} \;sr^{-1}}
\end{equation}
with an uncertainty of $\sim 40\%$ due to the experimental error on the
primary CR spectrum and composition and on theoretical models of
hadronic interactions. Notice that the background flux also depends on
zenith angle due to the different thickness of atmosphere to be crossed by
cosmic rays coming from different directions. This dependence is included
in the thickness of the shaded region by averaging on all possible
neutrinos' incoming directions (see \cite{barr06} for a detailed
description of the mean uncertainties in the neutrino flux determination).
Notice that for atmospheric neutrinos we neglect the contribution of
oscillations, which is relevant only for energies $\lesssim 10$ GeV
\cite[\S6.1]{lip06}.

With respect to Eq.~(\ref{eq:nu_atm}) the atmospheric background shown in
Fig.~\ref{fig:f2} is rescaled to a solid angle corresponding to a cone of
semi-aperture 0.5 degrees. The motivation for choosing this value of
the angle is twofold: it represents a reasonable estimate for the angular resolution
of a neutrino telescope at these energies, and it also corresponds to
approximately the angular size of the shell of RX~J1713.7-3946. A smaller value
of the aperture angle would imply a better chance to detect the
signal than estimated below.

In the following we specialize our predictions to the case of a $km^3$
neutrino telescope, for which the neutrino detection occurs by measuring
the Cherenkov light from $\nu$-induced muons produced by charged-current
interactions of $\nu_\mu$ and $\bar\nu_\mu$ in the matter just below the
detector.  For the computation of the production rates of $\mu$ and
$\bar\mu$, we follow the method of Ref.  \cite{cos05}.

Since neutrinos can only be detected when the source is below the horizon
of the detector, we introduce an average live-time of the source, in the
form of a parameter $f_{\rm liv}$, representing the fraction of time during
which this condition is satisfied. For a neutrino telescope as ANTARES and
RX~J1713.7-3946 as the source, one can estimate $f_{\rm liv}\approx 78\%$
\cite{cos05}.
On the other hand, absorption of neutrinos while crossing the Earth leads
to an energy dependent reduction of the detected flux. In fact the Earth
becomes opaque to neutrinos at energies $E_\nu \gtrsim 1$ PeV (when
$\sigma$ becomes larger than $10^{-33}$ cm$^2$). For neutrino energies of
$\sim 100$ TeV (10 TeV) the signal is reduced by about $20\%$ ($5\%$). 

Taking into account these two effects, the number of muons with energy
$E_\mu>E_{th}$, crossing an area $A$ during the observation time $T$ can be
written as:
\begin{eqnarray} \label{eq:mu_number}
N_{\mu}= f_{\rm liv} A\; T \int_{E_{th}}^\infty dE_\nu \,
 \Phi_\nu(E_\nu) \, Y_\mu(E_\nu,E_{th}) \times  \nn \\
 \left[1-\bar a_{\nu}(E_\nu)\right] \,, \hspace{1.8cm}
\end{eqnarray}
with a similar equation holding for $N_{\bar\mu}$. In Eq.
(\ref{eq:mu_number}), $\bar a_{\nu}(E_\nu)$ is the mean coefficient for
neutrino absorption through the Earth, including only charged-current
interactions and computed for a fixed Earth thickness resulting from
averaging over the observation time \cite[]{cos05}. $Y_\mu(E_\nu,E_{th})$
is the muon yield, namely the probability that a neutrino with energy
$E_\nu$ produces a muon with energy $E_\mu>E_{th}$ that crosses the
detector area.  This is obtained by integration over the muon energy,
$E_\mu$, of the neutrino interaction cross section multiplied by the muon
range,
$R(E_\mu,E_{th})$:
\begin{eqnarray} \label{eq:yield}
Y_{\mu}= \rho_{\rm H_2 O} \, N_A \int_{E_{th}}^{E_\nu} dE_\mu \,
 \frac{d\sigma_{cc}}{dE_\mu}(E_\nu,E_\mu) \, R(E_\mu,E_{th}) \,.
\end{eqnarray}
Here $\rho_{\rm H_2 O}$ is the water density and $N_A$ is Avogadro's
number. Both the muon range in water and the neutrino interaction
probability are taken from \cite{cos05}. The latter is calculated by using
the deep inelastic scattering formula for the charged-current
cross-section, $\sigma_{cc}$, and the distribution functions of partons as
calculated by \cite{mar07}. 

\begin{figure}
\begin{center}
\includegraphics[angle=0,scale=1]{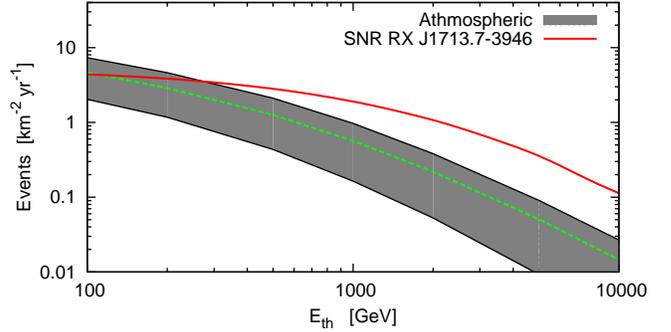}
\caption{Neutrino event rate per year from RX~J1713.7-3946 (\textit{solid
line}) as a function of muon energy threshold for a neutrino telescope
with 1 $km^2$ effective area. The dashed line shows the event rate
expected from atmospheric neutrinos integrated in a cone of semi-aperture
$0.5^\circ$.}
\label{fig:f3}
\end{center}
\end{figure}

In Fig.~\ref{fig:f3} the solid line shows the resulting $N_\mu+N_{\bar
\mu}$ for $A= 1\, {\rm km}^2$ and $T=1$ yr as a function of the energy
threshold. For $E_{th}= 50$ GeV the number of events per year is 4.7,
compatible with the findings of \cite{vis06}. 
The shaded region represents the muon background produced by atmospheric neutrinos, which can be computed as a function of the energy threshold, $E_{th}$, by substituting Eq.~(\ref{eq:nu_atm}) into Eq.~(\ref{eq:mu_number}). The result is normalized to an aperture angle of $0.5^\circ$, for the reasons explained above. The uncertainties represented by the shaded region have the same origin
as for Fig.~\ref{fig:f2}.

The spectrum of neutrinos from RX~J1713.7-3946 dominates the atmospheric background
at energies above $\sim 300$ GeV. In Table \ref{tab} we report the time (in
years) of observation needed to obtain a signal with significance level of
3$\sigma$, assuming a simple Poisson distribution of events and a unit
detection efficiency. One can see that if RX~J1713.7-3946 is indeed a hadronic
accelerator a signal at the 3$\sigma$ level could be seen in about 2 years of
observation at energies above $500$ GeV. 

\begin{table}
\caption{\label{tab}Comparison between the expected signal and
the atmospheric neutrino background. The last column shows the observation
time (in years) required to obtain a signal from RX~J1713.7-3946 with a
significance level of $3\sigma$. }
\begin{center}
\begin{tabular}{cccc}
\hline
$E_{th}$(GeV) & $N_{\mu+\bar\mu}$ & $N_{\mu+\bar\mu}^{\rm atm}$ &
yrs($3\sigma$)\\
\hline
\hline
100  & 4.4 & 4.7   & 2.16 \\
500  & 2.8 & 1.3   & 1.44 \\
1000 & 1.9 & 0.57  & 1.41 \\
\hline
\end{tabular}
\end{center}
\end{table}

\section{Conclusions}
\label{sec:conc}

In this paper we described the impact of non linear diffusive shock
acceleration for the gamma ray and neutrino production in SNRs. The
intricate connection between particle acceleration, magnetic field
amplification, dynamical reaction of the particles and magnetic field, and
the radiation produced by the accelerated particles (electrons and protons)
in principle allow, if taken at face value, to impose strong constraints on
the ability of SNRs to accelerate cosmic rays. An instance of how to use
this powerful tool was provided in \cite{morlino08}, where particle
acceleration was described in terms of the non linear theory of Refs.
\cite{amato05,amato06}. The magnetic field amplification was assumed to be
due to resonant streaming instability and both the dynamical reaction of
the accelerated particles and of the amplified field were taken into
account. The basic conclusion reached in \cite{morlino08} is that the
hadronic interpretation of the HESS data automatically leads to a
downstream magnetic field which is in agreement with that inferred from X-ray
observations if the rims of non-thermal emission are interpreted as
the result of synchrotron losses. The non-thermal X-ray spectrum, as
measured by Suzaku \cite{suzaku}, was also reproduced with
unprecedented accuracy. On the other hand, a satisfactory fit to the
data within a leptonic scenario could only be achieved by considerably
reducing the particle injection efficiency, through fine-tuning of the
only free parameter of our calculations, $\xi$, as defined in \S
\ref{sec:max_energy}. Moreover, in order for the leptonic model to fit
HESS data at all energies one is forced to require the presence of a
diffuse background of infrared light exceeding that observed in the
interstellar medium by more than a factor 20. These requirements
become less stringent if one decides not to include in the fit the
highest energy HESS data points, which strongly constrain the maximum 
energy of the radiating electrons. 

Since the discrimination between the two models, hadronic and leptonic, is
based on such a tricky region from the observational point of view, we
decided to address here the issue of how well the photon energies need to
be reconstructed at the telescope in order to tell the difference between
the predictions of a leptonic and a hadronic model. The issue is of
particular importance since the different predictions are of relevance in the
energy region where the particle spectra (and the gamma ray spectrum) are
already sharply falling and a small statistical uncertainty in the energy
determination may have sizable implications on the shape of the observed
spectrum. Our calculations show that a statistical uncertainty in the
energy determination larger than about 30\% would inhibit the possibility
to discriminate between a hadronic and a leptonic interpretation of TeV
data. Of course a systematic uncertainty would strengthen the problem.
These points need to be taken into account for the design of future
telescopes, such as CTA. Needless to say that the observation of TeV
emitting SNRs with the Fermi gamma ray telescope will definitely contribute
to settle the debate. This should certainly be the case for RX~J1713.7-3946
(see also \cite{morlino08}). 

In the absence of clear multifrequency evidence however, the smoking gun
that SNRs are efficient cosmic ray accelerators can only be provided by the
unambiguous detection of neutrinos. Here we used the same non linear theory
of particle acceleration to infer the number of neutrino induced muons in a
$km^{3}$ neutrino telescope. The flux is compared with the appropriate
background of atmospheric neutrinos. At the distance of RX~J1713.7-3946 and
assuming a hadronic interpretation of HESS data, we predict that a
$3\sigma$ detection should be achieved by $km^{3}$ neutrino telescope at
the Antares' location (which is in the right hemisphere to detect
RX~J1713.7-3946) in about 2 years of observation. 

\section*{Acknowledgments}
We are grateful to Damiano Caprioli and Gamil Cassam-Chenai for continuous
discussion and to Teresa Montaruli for reading the manuscript and providing
useful comments. This work was partially supported by MIUR (under grant
PRIN-2006) and by ASI through contract ASI-INAF I/088/06/0.

\end{document}